\documentclass[11pt,tightenlines,nofootinbib]{revtex4}
\usepackage{amsmath,amscd,amssymb}
\usepackage{epsfig,color,eso-pic}
\usepackage{graphics,mathtools}
\usepackage{diagbox,bigints}

\newcommand{\imu}{{\rm i}}

\def\sech{\mathop{\rm sech}\nolimits}
\def\csch{\mathop{\rm csch}\nolimits}
\def\arctanh{\mathop{\rm arctanh}\nolimits}

\begin{document}

\title{Quantum contribution to domain wall tension from spectral methods}

\author{N. Graham$^{a)}$, H. Weigel$^{b)}$}

\affiliation{
$^{a)}$Department of Physics, Middlebury College
Middlebury, VT 05753, USA\\
$^{b)}$Institute for Theoretical Physics, Physics Department,
Stellenbosch University, Matieland 7602, South Africa}

\begin{abstract}
In field theory, domain walls are constructed by embedding localized field 
configurations varying in
one space dimension, such as the $\phi^4$ kink, in two or three space 
dimensions. At the classical level, the kink mass straightforwardly turns into 
the energy per unit length or area, known as the domain wall tension. The quantum 
contribution to the tension is more difficult to compute, because the quantum 
fluctuations about the domain wall in the additional coordinates must be included. 
We show that spectral methods, making use of scattering data for the 
interaction of quantum fluctuations with the domain wall background, are an 
efficient way to compute the leading quantum correction to the domain wall 
tension. In particular we demonstrate that within this approach it is straightforward
to pass from one renormalization scheme to another.
\end{abstract}

\maketitle

\section{Introduction}

Non-linear field theories in one space dimension with degenerate translationally
invariant vacua typically allow for soliton (or solitary wave) solutions that mediate 
between various of these vacua as the spatial coordinate varies
from negative to positive spatial infinity \cite{Ra82}.
When such solitons are embedded in higher dimensions, they turn into domain walls that
separate domains with different physical properties. In cosmology 
there may be domains with different vacuum expectation values of scalar fields like 
the Higgs boson \cite{Zeldovich:1974uw,Kibble:1976sj}. Electric or magnetic materials 
may have regions with different polarization and/or magnetization 
\cite{PhysRevB.2.754,WOS:000574519100002}, while in condensed matter or statistical physics, 
the domain walls may separate regimes of different phases \cite{Salomaa:1988zz}.
For the many facets of domain walls in string theory, see {\it e.g.\@} 
chapter IV in Ref.~\cite{Tong:2005un}.
The soliton energy becomes the energy per unit length or area when embedded in two or
three space dimensions, respectively. These densities are frequently called tensions.
Like the soliton energy, the tension has classical and quantum contributions.

The domain wall problem in cosmology is a particular motivation to compute quantum corrections 
to the tension. If the tension exceeds a certain limit, domain walls will dominate the Universe's 
energy and cause significant anisotropies \cite{Vilenkin:1984ib}. Taking the mass of the scalar 
field fixed, the classical tension is proportional to the square of the vacuum expectation value 
of the scalar field and thus the domain wall problem sets an upper bound for this expectation 
value. The leading quantum correction to the tension only depends on the mass, so this correction 
will alter the bound.  In particular, if the correction is negative, as is typical for kinks 
\cite{Dashen:1974cj}, this bound will be increased. 

Here we will explore the vacuum polarization energy (VPE),
the leading (one-loop) quantum correction, for $n$ transverse
coordinates. This approach will also allow us to regularize the
ultraviolet divergent components by analytic continuation in 
$n$, and then renormalize them via standard counterterms. We will set up a
general approach, which we will then apply to the kink and sine-Gordon solitons.

There have been previous studies of tensions of domain walls constructed from the kink and
sine-Gordon solitons \cite{Rebhan:2002uk,Evslin:2024wwu}. Those
calculations involve multiplicative renormalizations of the classical soliton mass. This approach 
may fail when counterterm structures do not match the components of the classical mass, as is 
the case for the renormalization of the vacuum expectation value of the scalar field in the 
kink model. Our main focus will therefore be to show that spectral methods 
\cite{Graham:2009zz,Graham:2022rqk} can compute the tension in an efficient, constructive 
and transparent manner, precisely implementing  various renormalization conditions.
In this way we will be able to examine the causes of the discrepancies in the earlier
results, not only in magnitude but more importantly in sign. Since
Ref. \cite{Evslin:2024wwu} has attributed this difference to the
chosen renormalization scheme, it is important to analyze the VPE for
domain walls using an approach that easily links different
renormalization schemes.

In Sec.~\ref{sec:models} we will briefly introduce the classical soliton
models, and in Sec.~\ref{sec:interface} we
introduce the interface formalism \cite{Graham:2001dy} to compute the VPE.
In particular, we will consider the regularization prescription needed for finite results when $n=2$, 
and also apply it to $n=0,1$ as a consistency check. In Sec.~\ref{sec:renorm_kink} we will consider 
physically motivated renormalization schemes and compare to Refs. \cite{Rebhan:2002uk,Evslin:2024wwu}. 
For the particular models investigated here, analytic results are available. In Sec.~\ref{sec:dimreg} 
we will use these results to express the VPE renormalized at an arbitrary scale, and we will give 
concluding remarks in Sec.~\ref{sec:conclusion}.

\section{The models}
\label{sec:models}

We start from scalar models defined by the Lagrangian density
\begin{equation}
{\cal L} = \frac{1}{2} \dot \phi^2 - \frac{1}{2} \phi^{\prime2} - U(\phi)
\label{eq:lag}\end{equation}
in $D=1+1$ spacetime dimensions, with dot and prime denoting time and
space derivatives of $\phi$, respectively. The field potentials are
\begin{equation}
U_{\rm sG}=\frac{\mu^4}{\lambda}
\left[1-\cos\left(\frac{\sqrt{\lambda}\phi}{\mu}\right)\right]
\qquad {\rm and}\qquad 
U_{\rm kink}=\frac{\lambda}{8}\left[\phi^2-\frac{\mu^2}{\lambda}\right]^2
\label{eq:fieldpot}\end{equation}
for the sine-Gordon and kink models. Here $\mu$ is a mass parameter
and $\lambda$ is the interaction strength of the (classical)
four-point function. These models have static, localized solutions \cite{Ra82}
\begin{equation}
\phi_{\rm cl,sG}(x)=\frac{4\mu}{\sqrt{\lambda}}{\rm arctan}\left({\rm e}^{-\mu x}\right)
\qquad {\rm and}\qquad
\phi_{\rm cl,kink}(x)=\frac{\mu}{\sqrt{\lambda}}\tanh\left(\frac{\mu x}{2}\right)
\label{eq:soliton}\end{equation}
that we will call solitons, even though strictly speaking 
$\phi_{\rm cl,kink}(x)$ 
is only a solitary wave. These solitons have classical energies
\begin{equation}
E_{\rm cl,sG}=8\frac{\mu^3}{\lambda} 
\qquad {\rm and}\qquad
E_{\rm cl,kink}=\frac{2}{3}\frac{\mu^3}{\lambda}\,.
\label{eq:ecl}\end{equation}
We will then embed these solitons into $D=(n+1)+1$ dimensions with $n=0,1,2$, such that the configurations
are translationally invariant in the $n$ additional coordinates. The energies in Eq.~(\ref{eq:ecl}) then 
turn into energies per unit length ($n=1$) and area ($n=2$). Note that $\lambda$ has canonical energy 
dimension~$2-n$.

\section{Interface formalism with two subtractions}
\label{sec:interface}

Expanding the fields around the soliton as $\phi(t,x)=\phi_{\rm cl}(x)+\eta_k(x){\rm e}^{-\imu \omega t}$
with $\omega=\omega(k)=\sqrt{k^2+\mu^2}$ turns the field equations into a scattering problem for the 
fluctuation $\eta_k(x)$. The respective potentials are
\begin{equation}
V_{\rm sG}(x)=-2\mu^2\sech^2 \mu x \qquad {\rm and}\qquad
V_{\rm kink}(x)=-\frac{3\mu^2}{2}\sech^2\frac{\mu x}{2}\,.
\label{eq:pot}\end{equation}
From that scattering problem we determine the bound state solutions with energy eigenvalues
$\omega_j=\sqrt{\mu^2-\kappa_j^2}$ and phase shifts $\delta_{\pm}(k)$ in the symmetric and anti-symmetric
channels, which decouple because the potentials are invariant under $x\,\leftrightarrow\,-x$.
The total phase shift is $\delta(k)=\delta_{+}(k)+\delta_{-}(k)$ and
we denote the $j^{\rm th}$ order of its Born
expansion (power series in $V$) as $\delta_j(k)$.

The vacuum polarization energy (VPE) is the renormalized sum of the differences between the 
zero-point energies of the fluctuations in the presence and absence of the soliton. This sum 
contains bound and scattering state contributions. The latter is computed as the continuum 
integral over the energies $\omega(k)$ weighted by the change of the density of states
induced by the soliton. The central idea of spectral methods is to write this change
as the momentum derivative of the total phase shift \cite{Faulkner:1977aa}. In the 
$n$ trivial transverse coordinates the wavefunctions are simple plane waves. Their
momenta are integrated over in dimensional regularization, as given by 
 the bosonic case of Eq.~(7) in Ref.\ \cite{Graham:2001dy} for $m=1$,
\begin{align}
{\cal E}^{(n)}
&=-\frac{\Gamma\left(-\frac{n+1}{2}\right)}{2(4\pi)^{\frac{n+1}{2}}}\left[
\sum_j^{\rm b.s.}\left(\omega_j^{n+1}-\mu^{n+1}+\frac{n+1}{2}\kappa_j^2\mu^{n-1}\right)\right. 
\nonumber \\ &\hspace{1cm}\left.
+\int_0^\infty \frac{dk}{\pi}\left(\omega^{n+1}(k)-\mu^{n+1}-\frac{n+1}{2}k^2\mu^{n-1}\right)
\frac{d}{dk}\left(\delta(k)-\delta_1(k)-\delta_2(k)\right)\right]
\nonumber \\ &\hspace{1cm}
+E_{\rm FD}+E_{\rm CT}\,.
\label{eq:IF1}\end{align}
In what follows we will refer to $E_{\rm FD}+E_{\rm CT}$ as the perturbative part of the VPE.
The two subtractions from $\omega^{n+1}(k)$ are identities from (generalizations of) Levinson's
theorem and avoid infrared singularities at $n=0$ and anomalous contributions at $n=1$ \cite{Graham:2001iv}.
The second subtraction, in particular, causes the residue of the pole at $n=1$ to
vanish. The associated sum rule is \cite{Puff:1975zz}
$$
\int_0^\infty \frac{dk}{\pi}k^2\frac{d}{dk}\left(\delta(k)-\delta_1(k)\right)
=\sum_j^{\rm b.s.} \kappa_j^2\,.
$$
The Born subtractions render the momentum integral finite and are added back as Feynman diagrams 
yielding $E_{\rm FD}$. Together with the counterterm contribution, $E_{\rm CT}$, they make a finite 
contribution to the VPE.  Eq.~(\ref{eq:IF1}) has been numerically verified in Ref. \cite{Graham:2001iv} 
for potentials like those in Eq.~(\ref{eq:pot}). However, we want to employ the imaginary momentum
formulation, which has since then been observed to be more efficient in many aspects \cite{Graham:2022rqk}. 
This formulation starts from recognizing that, for real non-negative momenta $k$, the phase shift is 
the negative phase of the Jost function $F(k)$. The phase of the Jost function is odd for real $k$ while 
the magnitude is even; hence $\delta(k)=\frac{\imu}{2}\left[\ln F(k)-\ln F(-k)\right]$.
Most importantly, the Jost function is analytic for ${\sf Im}(k)\ge0$ with simple zeros at 
$k=\imu\kappa_j$ \cite{Newton:1982qc}. Thus the logarithmic derivative has poles with unit residue,
which in the contour integral exactly cancel the explicit bound state contribution 
in Eq.~(\ref{eq:IF1}) \cite{Bordag:1994jz}
Finally, the Born subtractions ensure that the integral along the 
semi-circle at $|k|\to\infty$ vanishes. The only contributions arise
from non-analytic pieces in Eq.~(\ref{eq:IF1}) contained in
$\omega^{\frac{n+1}{2}}(k)$, and thus it is sufficient to consider
the integral in
\begin{equation}
\mathcal{E}^{(n)} =
-\frac{\Gamma\left(-\frac{n+1}{2}\right)}{2(4\pi)^{\frac{n+1}{2}}}
\int_{-\infty}^\infty dk\,\left(k^2+m^2\right)^{\frac{n+1}{2}}
\frac{\imu}{2}\frac{d}{dk}\left[\ln F(k)\right]_p+E_{\rm FD}+E_{\rm
  CT} + \ldots
\label{eq:IF2}\end{equation}
where the ellipsis denotes bound state contributions, which eventually will cancel 
with corresponding poles. Here the subscript indicates the subtractions of 
the first $p$ terms of the Born expansion. Also, we did not write pieces that 
are analytic for ${\sf Im}(k)\ge0$ as they do not contribute to the contour 
integral. For $n=0$ and $n=2$, the $\Gamma$-function is regular
and the relevant discontinuities for  $k=\imu t$ with $t>\mu$ are
\begin{equation}
\left[\left(\imu t+0^{+}\right)^2+\mu^2\right]^\frac{n+1}{2}
-\left[\left(\imu t-0^{+}\right)^2+\mu^2\right]^\frac{n+1}{2}
=(-1)^{n}\imu\left[t^2-\mu^2\right]^\frac{n+1}{2}\,.
\label{eq:discont1}\end{equation}
For $n\approx1$ we encounter a pole 
$\Gamma\left(-\frac{n+1}{2}\right)\approx\frac{2}{n-1}+\ldots$. Then
\begin{equation}
\Gamma\left(-\frac{n+1}{2}\right)\omega^{\frac{n+1}{2}}
=\left[\frac{2}{n-1}+\ldots\right]
\left[\omega^2\left(1+\frac{n-1}{2}\ln\frac{\omega^2}{\mu^2}\right)\right]\,.
\label{eq:discont2}\end{equation}
We have introduced $\mu$ as the scale in the logarithm for convenience. It
is arbitrary because it adds an analytic piece in the contour integral\footnote{It
also vanishes by the sum rule for the scattering data. This is the same argument
because the sum rule arises from $F(k)$ being analytic.}. The only non-zero
contribution to the contour integral then arises from the discontinuity 
\begin{equation}
\ln\left[\left(\imu \frac{t}{\mu}+\epsilon\right)^2+1\right]
-\ln\left[\left(\imu \frac{t}{\mu}-\epsilon\right)^2+1\right]=2\pi\imu\,.
\label{eq:discont3}\end{equation}
Putting pieces together we find, with $p=2$,
\begin{align}
\mathcal{E}^{(0)}&=\int_0^\infty\frac{d\tau}{2\pi}\,
\left[\nu(t)-\nu_1(t)-\nu_2(t)\right]_{t=\sqrt{\tau^2+\mu^2}}
+E_{\rm FD}+E_{\rm CT}\,,\cr
\mathcal{E}^{(1)}&=\int_0^\infty\frac{d\tau}{4\pi}\,\tau
\left[\nu(t)-\nu_1(t)-\nu_2(t)\right]_{t=\sqrt{\tau^2+\mu^2}}
+E_{\rm FD}+E_{\rm CT}\,,\cr
\mathcal{E}^{(2)}&=\int_0^\infty\frac{d\tau}{4\pi^2}\,\tau^2
\left[\nu(t)-\nu_1(t)-\nu_2(t)\right]_{t=\sqrt{\tau^2+\mu^2}}
+E_{\rm FD}+E_{\rm CT}\,,
\label{eq:intf1}
\end{align}
where $\nu(t)=\ln F(\imu t)$ and the subscripts on $\nu$ denote
the order of its Born series. Since the potentials, Eq.~(\ref{eq:pot}),
do not contain the coupling constant~$\lambda$, the mass parameter
$\mu$ sets the scale for the $\tau$-integrals.

For $n=0$ and $n=1$, only the tadpole diagram is ultraviolet divergent and only 
the $\mathcal{O}(V)$ subtractions is needed. Then the no-tadpole (NT)
renormalization condition gives $\left(E_{\rm FD} + E_{\rm CT}
\right)_{\rm NT} =0$ and the VPE
becomes
\begin{equation}
\mathcal{E}_{\rm NT}^{(0)}=\int_0^\infty\frac{d\tau}{2\pi}\,
\left[\nu(t)-\nu_1(t)\right]_{t=\sqrt{\tau^2+\mu^2}}
\qquad {\rm and}\qquad
\mathcal{E}_{\rm NT}^{(1)}=\int_0^\infty\frac{d\tau}{4\pi}\,\tau
\left[\nu(t)-\nu_1(t)\right]_{t=\sqrt{\tau^2+\mu^2}}\,.
\label{eq:intf2}\end{equation}

For the sine-Gordon and kink models we have analytic expressions for $\nu$ 
and $\nu_1$ (but not $\nu_2$), {\it cf.} Sec. \ref{sec:dimreg}.
We can then use Eq.~(\ref{eq:intf2}) for $n=0$ and $n=1$. In the former case
we get the well-known quantum corrections \cite{Ra82}
\begin{equation}
\mathcal{E}_{\rm NT,sG}^{(0)}=-\frac{\mu}{\pi}\approx-0.31831\mu
\qquad {\rm and}\qquad
\mathcal{E}_{\rm   NT,kink}^{(0)}=
\mu\left(\frac{1}{4\sqrt{3}}-\frac{3}{2\pi}\right)
\approx-0.33313\mu\,,
\label{eq:result0}\end{equation}
while in the latter case we obtain \cite{Jaimungal:1998hk}
\begin{equation}
\mathcal{E}_{\rm NT,sG}^{(1)}=-\frac{\mu^2}{4\pi}\approx-0.079577\mu^2
\qquad {\rm and}\qquad
\mathcal{E}_{\rm NT,kink}^{(1)}=\frac{3\mu^2}{32\pi}\left(\ln3-4\right)
\approx-0.086582\mu^2\,.
\label{eq:result1}\end{equation}

The numerical computation of $\nu(t)$ and in particular $\nu_2(t)$ starts from the Jost 
solution $f(k,x)$ to the wave equation for the fluctuations. This solution 
behaves asymptotically like an outgoing plane wave. Parameterizing $f(k,x)=g(k,x){\rm e}^{\imu kx}$
and continuing $k=\imu t$, the factor-function is subject to the wave equation
\begin{equation}
g^{\prime\prime}(\imu t,x)=2tg^\prime(\imu t,x) +V(x)g(\imu t,x)
\label{eq:jost1}\end{equation}
with the boundary condition $\lim_{x\to\infty}g(\imu t,x)=1$. For spatially 
symmetric potentials, the scatting solutions are linear combinations of $f(k,x)$ and 
$f(-k,x)$ such that either the wavefunction or its derivative vanish at $x=0$. The 
coefficients are the Jost functions $F_{-}(\mp k)$ and $F_{+}(\mp k)$,
respectively. We then get \cite{Graham:2002xq}
\begin{equation}
\nu(t)=\lim_{x\to0}\ln\left[g(\imu t,x)
\left(g(\imu t,x)-\frac{1}{t}g^\prime(\imu t,x)\right)\right]\,.
\label{eq:jost2}
\end{equation}
We expand $g=1+g_1+g_2+\ldots$, where $g_p$ is
$\mathcal{O}(V^p)$. These functions vanish at spatial infinity and
solve the differential equations
\begin{equation}
g_1^{\prime\prime}(\imu t,x)=2tg_1^\prime(\imu t,x) +V(x)
\qquad{\rm and}\qquad
g_2^{\prime\prime}(\imu t,x)=2tg_2^\prime(\imu t,x) +V(x)g_1(\imu t,x) \,,
\label{eq:jost3}\end{equation}
which lead to
\begin{align}
\nu_1(t)+\nu_2(t)&=\lim_{x\to0}
\left[2\left(g_1(\imu t,x)+g_2(\imu t,x)\right)
-\frac{1}{t}\left(g_1^\prime(\imu t,x)+g_2^\prime(\imu t,x)\right)
\right.\cr & \hspace{4cm} \left.
-\frac{1}{2}g^2_2(\imu t,x)
-\frac{1}{2}\left(g_2(\imu t,x)-\frac{1}{t}g_2^\prime(\imu t,x)\right)^2\right]\,.
\label{eq:jost4}\end{align}
This completes the scattering part of the calculation. The perturbative part is
most straightforwardly obtained from the one-loop part of the effective action
$\mathcal{A}_{\rm eff}\sim\frac{\imu}{2}{\rm Tr}{\rm Log}
\left[\partial^2+\mu^2-\imu\epsilon+V\right]$. The $\mathcal{O}(V)$
contribution is fully canceled in the no-tadpole scheme, while the
second-order contribution
\begin{equation}
\mathcal{A}_{\rm eff}=-\frac{\imu}{4}{\rm Tr}\left[
\left(\partial^2+\mu^2-\imu\epsilon\right)^{-1}V
\left(\partial^2+\mu^2-\imu\epsilon\right)^{-1}V\right]
+ \mathcal{O}(V^3)
\label{eq:aeff1}\end{equation}
contains the relevant piece to be added back in. Standard techniques
yield for $n=0,1$
\begin{equation}
\mathcal{A}_{\rm eff}=\frac{1}{2^n\cdot 16\pi}
\int \frac{d^{n+2}p}{(2\pi)^{n+2}}\int_0^1d\alpha\,\frac{|\widetilde{V}{(p)}|^2}
{\left[\mu^2-\alpha(1-\alpha)p^2\right]^{1-\frac{n}{2}}} + \mathcal{O}(V^3)
\,,
\label{eq:aeff2}\end{equation}
where $\widetilde{V}{(p)}$ is the Fourier transform of the potential in Eq.~(\ref{eq:jost1}).
For static potentials, only space-like Fourier momenta contribute and the Feynman 
parameter integrals can be computed without any further restrictions,
giving
\begin{equation}
E_{\rm FD}=-\frac{\mu^{n+1}}{2^n\cdot4\pi^2}\int_0^\infty ds\, v^2(s)I_n(s)\,,
\label{eq:efd1}\end{equation}
where $v(s)=\bigintsss_0^\infty dx\,\cos(sx)\,V(x)\Big|_{\mu=1}$,
\begin{equation}
I_0(s)=-2\frac{\ln\left[1+\frac{s^2}{2}-\frac{s}{2}\sqrt{4+s^2}\right]}{s\sqrt{4+s^2}}
\qquad{\rm and}\qquad
I_1(s)=\frac{2}{s}\,{\rm arctan}\left(\frac{s}{2}\right)\,.
\label{eq:efd2}\end{equation}
Numerical results from this formalism for the kink and sine-Gordon models are 
listed in Table \ref{tab:1}.
\begin{table}[b]
\centerline{
\begin{tabular}{l|cc}
$n=0$ & kink & sG \\
\hline
$\int d\tau\ldots$  & -0.21728 & -0.22210\\
$E_{\rm FD}$ & -0.11584 & -0.09621\\
$\mathcal{E}_{\rm NT}^{(0)}$       & -0.33312 & -0.31831
\end{tabular}
\hspace{2cm}
\begin{tabular}{l|cc}
$n=1$ & kink & sG \\
\hline
$\int d\tau\ldots$  & -0.027806& -0.029196\\
$E_{\rm FD}$ & -0.058776& -0.050381\\
$\mathcal{E}_{\rm NT}^{(1)}$       & -0.086582& -0.079576
\end{tabular}}
\caption{\label{tab:1}The VPE for $n=0$ (left) and $n=1$ (right)
numerically computed according to Eq.~(\ref{eq:intf1}) in the
no-tadpole (NT) scheme. The entry labeled
$\int d\tau\ldots$ denotes the integral contribution in Eqs.~(\ref{eq:intf1}), 
$E_{\rm FD}$ is from Eq.~(\ref{eq:efd1}) and $\mathcal{E}_{\rm NT}$
is their sum. For convenience we have set $\mu=1$.}
\end{table}
As expected, we find perfect agreement with the exact results in Eqs.~(\ref{eq:result0}) 
and~(\ref{eq:result1}). This verifies the VPE calculation with imaginary momenta and two 
subtractions, which will become compulsory for $n=2$. Not only are Eqs.~(\ref{eq:intf1})
much more compact than what can be extracted from Eq.~(\ref{eq:IF1}), they also
avoid the need to compute the bound state energies (although they are
known analytically for the
two examples considered). The formalism can be directly generalized to any 
spatially symmetric potential. If this symmetry is absent, one has to follow the
treatment of App.~B in Ref. \cite{Graham:2002xq}.

We have also numerically computed the VPE according to Eq.~(\ref{eq:intf2}) and indeed
obtained agreement with the analytic results listed in Eqs.~(\ref{eq:result0}) and~(\ref{eq:result1}). 
However, in that approach the numerical cut-off on the $\tau$ integral had taken an order of
magnitude larger, because $\nu(t)-\nu_1(t)$ approaches zero
significantly more slowly 
than $\nu(t)-\nu_1(t)-\nu_2(t)$ when $t\to\infty$.

Having established the two-subtraction formalism, we may now proceed to the case with
two transverse coordinates, $n\to2$, {\it i.e.} $D=3+1$. The momentum integral in 
Eq.~(\ref{eq:intf1}) only requires a slight modification of the $n=0$ and $n=1$
cases. The essential novelty is the second-order term of the effective action, 
Eq.~(\ref{eq:aeff2}), since the first-order term is still absent in the no-tadpole scheme.
In dimensional regularization the second-order term reads
\begin{align}
\mathcal{A}_{\rm eff}&=\frac{\Lambda^{2-n}}{4(4\pi)^{\frac{n}{2}+1}}
\Gamma\left(1-\frac{n}{2}\right)
\int \frac{d^{n+2}p}{(2\pi)^{n+2}}\int_0^1d\alpha\,\frac{|\widetilde{V}{(p)}|^2}
{\left[\mu^2-\alpha(1-\alpha)p^2\right]^{1-\frac{n}{2}}}
+\mathcal{O}(V^3) \cr
&=\frac{1}{64\pi^2}\int \frac{d^4p}{(2\pi)^4}|\widetilde{V}{(p)}|^2
\left[C_n-\int_0^1d\alpha\,\ln\left(1-\alpha(1-\alpha)\frac{p^2}{\mu^2}\right)\right]
+\mathcal{O}(V^3)
\label{eq:aeff3}\end{align}
for $n\,{\scriptstyle\lesssim}\,2$, where
$C_n=\frac{2}{2-n}-\gamma+\ln\frac{4\pi\Lambda^2}{\mu^2}$ and we have
dropped terms of order $n-2$ in the second equation.
We have introduced the scale $\Lambda$ such that the loop integral has the same dimensions as 
for $n=2$.  Since $\bigintsss\frac{d^4p}{(2\pi)^4}|\widetilde{V}{(p)}|^2=\bigintsss d^4x V^2(x)$ 
we can cancel the ultraviolet divergence in $C_n$ by adding a Lagrangian counterterm 
$\mathcal{L}_{\rm CT}\propto C_n  V^2(x)$, which is the $\overline{\rm MS}$ renormalization 
scheme.  For the kink model $V^2\propto (\phi^2-v^2)^2$, and this counterterm renormalizes the 
coupling constant~$\lambda$. For the sine-Gordon model, $V^2$ is not part of the original 
Lagrangian; the model is not renormalizable in $D=3+1$ dimensions since the only available 
non-derivative counterterm is proportional to $V$ \cite{deVega:1976sm}. As a matter of 
completeness we can nevertheless compute the correspondingly subtracted quantity by omitting 
the $C_n$ piece. For brevity and convenience we call this treatment of the sine-Gordon model
an $\overline{\rm MS}$ scheme as well. Carrying out the Feynman parameter integral 
and substituting the soliton profile yields
\begin{equation}
\left(E_{\rm FD}+E_{\rm CT}\right)_{\overline{\rm MS}}
=\frac{\mu^3}{8\pi^3}\int_0^\infty ds\, v^2(s)\left[I_2(s)-1\right]
\quad{\rm with}\quad I_2(s)=\frac{1}{s}\sqrt{4+s^2}\,{\rm arsinh}\left(\frac{s}{2}\right)
\label{eq:efd3}\end{equation}
for the perturbative part of the VPE per unit area. A consistency check is that the expression
in square brackets in Eq.~(\ref{eq:efd3}) vanishes at $s=0$ in the 
$\overline{\rm MS}$ scheme. Our numerical results are presented
in Table \ref{tab:2}.
\begin{table}[t]
\centerline{
\begin{tabular}{l|ccc}
& kink &~~~& sG \\
\hline
$\int d\tau\ldots$      & $-6.0403\times10^{-3}$ && $-6.5534\times10^{-3}$ \\
$E_{\rm FD}+E_{\rm CT}$ & $\hspace{2.5mm}0.2974\times10^{-3}$  && $\hspace{2.5mm}0.9246\times10^{-3}$ \\
$\mathcal{E}_{\overline{\rm MS}}^{(2)}$ & $-5.7429\times10^{-3}$ &&
$-5.6288\times10^{-3}$
\end{tabular}}
\caption{\label{tab:2}The VPE for $n=2$ numerically computed according to 
Eq.~(\ref{eq:intf1}), which implements the 
$\overline{\rm MS}$-no-tadpole scheme for the kink model.  For the
sine-Gordon model, which is not renormalizable in
this dimension, it implements the analogous subtractions. The
entry labeled $\int d\tau\ldots$ denotes the integral contribution in
Eqs.~(\ref{eq:intf1}), $E_{\rm FD}+E_{\rm CT}$ is from
Eq.~(\ref{eq:efd3}) and  $\mathcal{E}_{\overline{\rm MS}}^{(2)}$ is
their sum. For convenience we have set $\mu=1$.}
\end{table}
Within this renormalization scheme, the perturbative part of the VPE tends to
slightly reduce the magnitude of the quantum correction.

\section{Physically motivated schemes for kink}
\label{sec:renorm_kink}

In Ref.\ \cite{Rebhan:2002uk} the surface tensions for the kink were computed 
for several physically motivated renormalization schemes. Those authors were able to 
express both the phase shift and the counterterm coefficients in terms of  
hyper-geometric functions of the number of transverse dimensions, which is specific 
to the kink model.

Different schemes have different conditions on the coefficients in the counterterm 
Lagrangian, but the momentum integrals in Eq.~(\ref{eq:intf1}) are unaffected. This is
the great advantage of our approach: we only need to adjust $E_{\rm FD}+E_{\rm CT}$.
The authors of Ref.\ \cite{Rebhan:2002uk} consider four different sets of renormalization 
conditions that they label MR, OS, ORS, and ZM. All four implement a
no-tadpole scheme, but differ by the conditions on the two-point
function for the quantum fluctuations about the translationally
invariant vacuum.

The minimal renormalization (MR) is the pure no-tadpole scheme. It 
is only applicable for $n=0$ and $n=1$, and corresponds to our results in
Table \ref{tab:1}.

The general counterterm Lagrangian in the kink model is 
\begin{equation}
\mathcal{L}_{\rm CT}=\frac{c_0}{2}\partial_\mu\phi\partial^\mu\phi
+\frac{c_1}{8}\left(\phi^2-v^2\right)^2+\frac{c_2}{2}\left(\phi^2-v^2\right)\,,
\label{eq:lct}\end{equation}
with $v^2=\frac{\mu^2}{\lambda}$. Note that $c_0$ is finite at
one-loop order for $n\leq 2$. In the no-tadpole scheme, we ignore both the
$\mathcal{O}(V)$ term in the expansion of $\mathcal{A}_{\rm eff}$ and the $c_2$ term 
above. To determine $c_0$ and $c_1$, the authors of \cite{Rebhan:2002uk} imposed conditions 
on the polarization functions for fluctuation about the translationally invariant vacuum,
$\phi(x)=v+h(x)$. With $V=3\mu\sqrt{\lambda}h+\ldots$ we need the
$\mathcal{O}(V^2)$ quantum contribution, since this term contains the only
$\mathcal{O}(h^2)$ contribution in the
no-tadpole scheme. We renormalize by adding the $\mathcal{O}(h^2)$ pieces
of $\bigintsss d^Dx \mathcal{L}_{\rm CT}$,
\begin{equation} 
\mathcal{A}_{\rm ren,eff}=\mu^2\lambda\int\frac{d^{n+2}p}{(2\pi)^{n+2}}
|\widetilde{h}(p)|^2\Pi_h(p^2) + \mathcal{O}(h^3) \,,
\label{eq:renorm1}\end{equation}
with the polarization function
\begin{equation}
\Pi_h(p^2)=-\frac{9\imu}{4}\int\frac{d^{n+2}l}{(2\pi)^{n+2}}
\int_0^1\frac{d\alpha}{\left[l^2-\mu^2+\imu\epsilon+\alpha(1-\alpha)p^2\right]^2}
+\frac{c_0}{2\mu^2\lambda}p^2+\frac{c_1}{2\lambda^2}\,,
\label{eq:renorm2}\end{equation}
where the $h$ subscript denotes the insertion of this field at
the vertices of the two-point function.
The on-shell scheme (OS) sets $c_0=0$ and determines $c_1$ from
$\Pi_h(\mu^2)=0$. In our formulation, this corresponds to 
\begin{equation}
\left(E_{\rm FD}+E_{\rm CT}\right)_{\rm OS}
=K_n\mu^{n+1}\int_0^\infty ds\, v^2(s)\left[I_n(s)-I_n\right]\,,
\label{eq:renorm3}\end{equation}
with the coefficients $K_0=-\frac{1}{4\pi}$, $K_1=-\frac{1}{8\pi^2}$ and 
$K_2=\frac{1}{8\pi^3}$ read off from Eqs.~(\ref{eq:efd2}) and~(\ref{eq:efd3}).
The subtractions are  $I_0=\frac{2\pi}{3\sqrt{3}}$, $I_1=\ln3$
and $I_2=\frac{\pi}{2\sqrt{3}}$. By construction, $I_n(\imu)=I_n$.

The on-shell-residue (OSR) scheme augments the OS conditions
by requiring that the residue of the propagator does not have any 
quantum correction. This condition changes the above $c_1$ by
$-c_0\lambda$, and extracts $c_0$ from
$\frac{\partial \Pi_h(\mu^2)}{\partial \mu^2}=0$. In total, the constants
added to the above are
\begin{equation}
\left(E_{\rm FD}+E_{\rm CT}\right)_{\rm OSR}=
\left(E_{\rm FD}+E_{\rm CT}\right)_{\rm OS} + 
\Delta\widetilde{\mathcal{E}}^{(n)} \,,
\label{eq:renorm4}\end{equation}
where $\Delta\widetilde{\mathcal{E}}^{(0)}=\frac{\sqrt{3}\pi-9}{18\pi} \mu$, 
$\Delta\widetilde{\mathcal{E}}^{(1)}=\frac{1}{16\pi}\left(3\ln3-4\right)\mu^2$,
and $\Delta\widetilde{\mathcal{E}}^{(2)}
=\frac{3\sqrt{3}-2\pi}{16\sqrt{3}\pi}\mu^3$.

Finally, the zero mass (ZM) condition requires that
$\Pi_h(p^2)=\sum_{l=2}a_l(p^2)^l$. This condition
determines $c_1$ and $c_0$ from $\Pi_h(0)$ and $\frac{\partial
  \Pi_h(p^2)}{\partial p^2}\Big|_{p^2=0}$,
respectively. The perturbative part of the VPE is then
\begin{equation}
\left(E_{\rm FD}+E_{\rm CT}\right)_{\rm ZM}
=K_n\mu^{n+1}\int_0^\infty ds\,
v^2(s)\left[I_n(s)-1\right]+
\Delta\widetilde{\mathcal{E}}^{(n)}_0 \,,
\label{eq:renorm5}\end{equation}
with $\Delta\widetilde{\mathcal{E}}^{(0)}_0
=-\frac{1}{16\pi}\mu$,
$\Delta\widetilde{\mathcal{E}}^{(1)}_0 = -\frac{1}{64\pi}\mu^2$ and
$\Delta\widetilde{\mathcal{E}}^{(2)}_0 = -\frac{1}{64\pi^2} \mu^3$.
It is also possible to derive an analytic expression for the integral
$\bigintsss_0^\infty ds\ v^2(s)$ and combine those pieces with 
$\Delta\widetilde{\mathcal{E}}^{(n)}$ or
$\Delta\widetilde{\mathcal{E}}^{(n)}_0$.  We will use this approach in
Sec.\ \ref{sec:dimreg}, but here we prefer the above formulation because
$I_n(\imu)=I_n$  and $I_n(0)=1$.
Our numerical results are listed in Table \ref{tab:3}.
\begin{table}[t]
\centerline{
\begin{tabular}{c|ccc}
$n$ & $\mathcal{E}_{\rm OS}^{(n)}$ & $\mathcal{E}_{\rm OSR}^{(n)}$ 
& $\mathcal{E}_{\rm ZM}^{(n)}$\\
\hline
$0$  & $-0.18878$ & $-0.25171$ & $-0.23365$\\
$1$  & $-2.1013\times10^{-2}$ & $-3.5022\times10^{-2}$ & $-3.1872\times10^{-2}$\\
$2$  &$-3.9742\times10^{-3}$ & $-7.9485\times10^{-3}$ & $-7.3260\times10^{-3}$
\end{tabular}}
\caption{\label{tab:3}Domain wall tension for various renormalization schemes
described in the text for $\mu=1$.}
\end{table}
In all cases considered we agree with the results presented in Ref.\ \cite{Rebhan:2002uk}.

The renormalization scheme of Ref.\ \cite{Evslin:2024wwu} only includes counterterms
that are compulsory for $n=2$:
\begin{equation}
\mathcal{L}_{\rm CT}=\frac{c_1}{8}\left(\phi^2-v^2\right)^2+
\frac{c_2}{2}\left(\phi^2-v^2\right)\,.
\label{eq:evslin1}\end{equation}
That scheme has a condition on the three-point function, so we expand the quantum 
correction to the effective action up to third order in $V$ to collect all contributions
$\mathcal{O}(h^3)$ that build the three-point function. With no contribution
linear in $h$ (since it would lead to a quantum correction for the VEV), this expansion yields
\begin{equation}
\mathcal{A}_{\rm ren,eff}\sim \int \frac{d^{n+2}p}{(2\pi)^{n+2}}\,|\widetilde{h}(p)|^2\Pi_h(p^2)
+\int\frac{d^{n+2}p}{(2\pi)^{n+2}}\int\frac{d^{n+2}q}{(2\pi)^{n+2}}\,
\widetilde{h}(p)\widetilde{h}(q)\widetilde{h}(-p-q)
\Gamma_3(p,q) + \mathcal{O}(h^4)\,.
\label{eq:evslin2}\end{equation}
Without the no-tadpole condition, the polarization function
\begin{align}
-\frac{3\mu^2\lambda}{4(4\pi)^{\frac{n}{2}+1}}\Gamma\left(-\frac{n}{2}\right)
\left(\mu^2\right)^{\frac{n}{2}-1}
+\frac{9\mu^2\lambda}{4(4\pi)^{\frac{n}{2}+1}}\Gamma\left(1-\frac{n}{2}\right)
\int_0^1 d\alpha \left[\mu^2-\alpha(1-\alpha)p^2\right]^{\frac{n}{2}-1}
+\frac{c_2}{2}+\frac{c_1}{2}\frac{\mu^2}{\lambda}
\label{eq:evslin3}\end{align}
has two ultraviolet divergent contributions, one from $\mathcal{O}(V)$ and another one
from $\mathcal{O}(V^2)$. The (amputated) three-point function is
\begin{align}
\Gamma_3(p,q) &=\frac{9}{4}\frac{\mu\lambda^{3/2}}{(4\pi)^{\frac{n}{2}+1}}
\Gamma\left(1-\frac{n}{2}\right)\int_0^1 d\alpha \left[\mu^2-\alpha(1-\alpha)p^2\right]^{\frac{n}{2}-1}\cr
&\hspace{0.5cm}-\frac{9}{2}\mu^3\sqrt{\lambda}^3\imu\int\frac{d^{n+2}l}{(2\pi)^{n+2}}
\frac{1}{l^2-\mu^2+\imu\epsilon} \frac{1}{(l-p)^2-\mu^2+\imu\epsilon}\frac{1}{(l-q)^2-\mu^2+\imu\epsilon}
+\frac{c_1}{2}\frac{\mu}{\sqrt{\lambda}}\,.\quad
\label{eq:evslin4}\end{align}
It has contributions from $\mathcal{O}(V^2)$ and from $\mathcal{O}(V^3)$.
In this language the renormalization conditions of Ref.\ \cite{Evslin:2024wwu} are
\begin{equation}
\Pi_h(\mu^2)=0 \qquad {\rm and}\qquad \Gamma_3(0,0)=0\,,
\label{eq:evslin5}\end{equation}
which we refer to as the ${\rm MG}_3$ scheme since it is specified via
the mass and the three-point Green's function.
In general, the three-point function is a complicated object to calculate \cite{tHooft:1978jhc};
fortunately not so for vanishing external momenta. Since there are two Born subtractions in Eq. (\ref{eq:intf1}),
we can read off the Feynman diagram contributions to be added back
from the quantum contribution to the effective action that is second order in $V$
\begin{align}
\mathcal{A}_{\rm ren,eff}&=\frac{\mu^{n-2}}{4}\frac{\Gamma\left(1-\frac{n}{2}\right)}{(4\pi)^{\frac{n}{2}+1}}
\int \frac{d^{n+2}p}{(2\pi)^D}|\widetilde{V}(p)|^2
\int_0^1d\alpha\left\{\left[1-\alpha(1-\alpha)\frac{p^2}{\mu^2}\right]^{\frac{n}{2}-1}-1\right\}\cr
&\hspace{1cm}+\int d^{n+2}x\left[\frac{\overline{c}_1}{8}\left(\phi^2-v^2\right)^2
+\frac{\overline{c}_2}{2}\left(\phi^2-v^2\right)\right] + \mathcal{O}(V^3)
\label{eq:evslin7}\end{align}
with
\begin{align}
\overline{c}_1&=\frac{9\lambda^2\mu^{n-2}}{2(4\pi)^{\frac{n}{2}+1}}\Gamma\left(2-\frac{n}{2}\right)
\qquad {\rm and}\cr
\overline{c}_2&=\frac{9}{2}\frac{\lambda\mu^{n}}{(4\pi)^{\frac{n}{2}+1}}
\left\{\Gamma\left(1-\frac{n}{2}\right)\left[1 -\int_0^1d\alpha\left(1-\alpha(1-\alpha)\right)^{\frac{n}{2}-1}\right]
-\Gamma\left(2-\frac{n}{2}\right)\right\}\,.
\label{eq:evslin8}\end{align}
The term with $\Gamma\left(1-\frac{n}{2}\right)$ in $\overline{c}_2$
cancels the Feynman parameter integral for $p^2=\mu^2$ in
Eq.~(\ref{eq:evslin7}) at $\mathcal{O}(h^2)$. For $n=0$ and
$n=1$, we may omit the ``1'' in both of
these terms, but we keep it to illustrate that both terms are finite as $n\,\to\,2$. The terms with 
$\Gamma\left(2-\frac{n}{2}\right)$ in $\overline{c}_1$ and $\overline{c}_2$ cancel the contributions 
$\mathcal{O}(h^2)$ in the counterterms, so we have implemented
$\Pi_h(\mu^2)=0$. Finally, we note that the
$\Gamma\left(2-\frac{n}{2}\right)$ term in $\overline{c}_1$ cancels the $h^3$ term originating from
$\mathcal{O}(V^3)$ in $\mathcal{A}_{\rm eff}$ for zero external momenta. Since the curly bracket 
in Eq.~(\ref{eq:evslin7}) vanishes for $p^2=0$, we indeed have $\Gamma_3(0,0)=0$.

The momentum integral in Eq.~(\ref{eq:evslin7}) leads to the Feynman diagrams that contribute 
to the VPE as in Eq. (\ref{eq:renorm3}) with $I_n=1$, while the finite counterterms give 
\begin{equation}
\left(\Delta E_{\rm CT}\right)_{{\rm MG}_3}
=-\frac{\overline{c}_1}{4}\int_0^\infty dx\,\left[v^2-\phi^2\right]^2
+\overline{c}_2\int_0^\infty dx\,\left[v^2-\phi^2\right]
=-\frac{\overline{c}_1}{3}\frac{\mu^3}{\lambda^2}+2\overline{c}_2\frac{\mu}{\lambda}\,.
\label{eq:DeltaEct}\end{equation}
Note that $\Delta E_{\rm CT}$ will have an overall factor of $\mu^{n+1}$
and no $\lambda$ dependence. Results are presented in Table~\ref{tab:evslin}. 
\begin{table}[ht]
\centerline{
\begin{tabular}{l|cccc|c}
$n$ & $\overline{c}_1$ & $\overline{c}_2$ & $\Delta E_{\rm CT}$ & 
$\mathcal{E}_{{\rm MG}_3}^{(n)}$ & Ref.\ \cite{Evslin:2024wwu}\cr
\hline
$0$ & $\frac{9}{8\pi}$ & $-\frac{\sqrt{3}}{4}$ & $-0.98539$ & $-1.19915$ & $0.38856$\\[1mm]
$1$ & $\frac{9}{32\pi}$ & $\frac{9}{32\pi}\left(1-2\ln(3)\right)$ & $-0.24420$ & $-0.27110$ & $0.12189$\\[1mm]
$2$ & $\frac{9}{32\pi^2}$ & $\frac{9}{32\pi^2}\left(\frac{\pi}{\sqrt{3}}-3\right)$ &
$-0.07710$ & $-0.08285$ & $0.04110$
\end{tabular}}
\caption{\label{tab:evslin}Results for the renormalization scheme of
Ref.\ \cite{Evslin:2024wwu} with $\mu=1$. The last column lists the
VPE prediction from that reference. Here $\overline{c}_1$ and
$\overline{c}_2$ are calculated with $\lambda=1$, while the $\lambda$
dependence cancels in the remaining quantities.}
\end{table}
Obviously there are significant discrepancies, both in sign and
magnitude, compared to the results of
Ref.\ \cite{Evslin:2024wwu}, which does not substitute the soliton into the counterterm Lagrangian. 
Rather the multiplicative renormalization of $\mu$ and $\lambda$
is substituted into the expression for the classical energy.  However, the 
(finite) $\overline{c}_2$-type counterterm is not part of the classical Lagrangian; it enters 
only via $v^2\,\longrightarrow\,v^2+\Delta v^2$ at $\mathcal{O}\left(\Delta v^2\right)$.
We note that Ref.\ \cite{Rebhan:2002uk} applies a similar multiplicative procedure for the finite 
wavefunction renormalization without explicitly introducing the $c_0$ type counterterm listed in
Eq. (\ref{eq:lct}). Doing so does not cause a problem in that case, however, because 
$\bigintsss dx\,\phi^{\prime2}(x)=2\bigintsss dx\,U(\phi(x))=\frac{1}{9\lambda}\bigintsss dx\,V^2(\phi(x))$ 
for a soliton solution according to Derrick's theorem \cite{Derrick:1964ww}. That is, for the soliton 
the spatial integral of the $c_0$-type counterterm is that of the $c_1$-type counterterm.

As our general calculation shows, the spectral method approach is
straightforwardly applicable even in cases where there is no analytic
expression for  the fluctuation
potential and/or the Jost function. For the kink and sine-Gordon
models, however, we can use exact results to avoid the need for numerical
simulations, as we will describe in the following section.

\section{General renormalization scheme in dimensional regularization}
\label{sec:dimreg}

In the preceding calculation, we have subtracted the full diagram
contributions and added them back in combination with renormalization
counterterms.  By using dimensional regularization, however, we can
shortcut this process by subtracting the counterterm directly.  (For
the tadpole graph, these subtractions are identical because the
diagram is local.)  This approach allows us to carry out the full
calculation analytically, at a general renormalization scale $M$, with
$M=\mu$ and $M=0$ corresponding to the OS and $\overline{\rm MS}$ or
ZM schemes above.

For general transverse dimension $n$, closing the contour for the
integral in Eq.~(\ref{eq:IF2}) and using
\begin{equation}
\Omega^{n+1} \left(\imu^{n+1} - (-\imu)^{n+1}\right) = 
2\imu \Omega^{n+1} \sin \frac{(n+1)\pi}{2} \,,
\end{equation}
where $\Omega = \sqrt{t^2 - \mu^2}$, along with
\begin{equation}
\sin \pi z  = -\frac{\pi}{\Gamma\left(z+1\right) \Gamma\left(-z\right)}
\end{equation}
yields \cite{Graham:2002yr}
\begin{equation}
\mathcal{E}_{\rm NT}^{(n)} = -\frac{1}{2 (4\pi)^{\frac{n+1}{2}}
\Gamma\left(\frac{n+3}{2}\right)} \int_\mu^\infty (t^2 - \mu^2)^{\frac{n+1}{2}}
\frac{\partial }{\partial t} \left[\nu(t) - \nu_1(t) \right] \, dt
\label{eq:VPE2002}
\end{equation}
for the VPE in the no-tadpole formulation with one subtraction, where $\nu(t)$ is given by 
Eq.~(\ref{eq:jost2}). As before, this expression denotes the VPE per
$L^n$, which 
is the generalized volume of the trivial coordinates.

We can write both scattering potentials in the general form
\begin{equation}
V(x) = - \frac{\ell+1}{\ell} \mu^2 \sech^2 \frac{\mu x}{\ell} \,,
\end{equation}
with $\ell=1$ for the sine-Gordon soliton and $\ell=2$ for the kink.
These are exactly solvable P\"oschl-Teller potentials, with
\begin{eqnarray}
g^{\rm sG}(k,x) &=& \frac{k + \imu \mu \tanh \mu x}{k+\imu \mu}
\cr
g^{\rm kink}(k,x) &=&  \frac{1}{k+\imu \mu}\frac{1}{k+\imu \frac{\mu}{2}}
\left(\frac{\mu^2}{4}+k^2+\frac{3}{2}\imu \mu k \tanh
\frac{\mu x}{2}-\frac{3}{4}\mu^2
\tanh^2\frac{\mu x}{2}\right) \,.
\end{eqnarray}

Similarly, the first Born approximation is given by \cite{Graham:2001iv}
\begin{equation}
\nu_1(t) = 2 g_1(\imu t,0) 
- \frac{1}{t} g_1^\prime(\imu t,0)
\end{equation}
with
\begin{align}
g_1(k,x) &=\frac{\imu}{2k} \int_x^\infty
\left(1-e^{2\imu k (y-x)}\right) V(y) \, dy \cr
&= \imu (\ell+1) \left[\ell \pi e^{-2 \imu k x}
\csch \frac{\ell k \pi}{\mu} - \frac{\mu}{k} 
{}_2 F_1 \left(1,\frac{\imu \ell k}{\mu},1+\frac{\imu \ell k}{\mu},
-e^{\frac{2 \mu x}{\ell}}\right) \right] \,.
\end{align}
Since $V(-x)=V(x)$ this gives
\begin{equation}
\nu_1(t) = \frac{\langle V \rangle}{2t} =-(\ell+1) \frac{\mu}{t}  \,,
\end{equation}
written in terms of the average value of the potential per unit length
or area,
\begin{equation}
\langle V(x) \rangle = \int_{-\infty}^{\infty} V(x) \, dx 
= -2 \mu (\ell+1) \,.
\end{equation}

For $n=0$ and $n=1$, subtracting the tadpole graph is sufficient to
renormalize the theory. For these cases we have the standard results
given in Eqs.~(\ref{eq:result0}) and~(\ref{eq:result1}).
The key observation for other renormalization schemes is that we can write 
the loop integral in the polarization function,
Eq.~(\ref{eq:renorm2}), for $n<2$ as \cite{Graham:2002yr} 
\begin{align}
\Pi_V(M^2)&=-\frac{\imu}{4} \int_0^1 d\alpha
\int \frac{dE}{2\pi} \frac{d^{n+1} q}{(2\pi)^{n+1}}
\frac{1}{(E^2 - q^2 - \mu^2 + M^2 \alpha (1-\alpha) + \imu \epsilon)^2} \cr
&= \frac{1}{2(4\pi)^{\frac{n+1}{2}} \Gamma\left(\frac{n+1}{2}\right)}
\int_0^\infty \frac{q^{n}}{\omega (4\omega^2-M^2)} \, dq 
\qquad {\rm with}\qquad \omega = \sqrt{q^2 + \mu^2}\cr
&= -\frac{1}{2(4\pi)^{\frac{n+1}{2}}\Gamma\left(\frac{n+3}{2}\right)}
\int_\mu^\infty (t^2 - \mu^2)^{\frac{n+1}{2}}
\frac{\partial}{\partial t}\left( \frac{1}{2t} \cdot \frac{1}{4t^2-M^2} \right) \, dt \,,
\label{eq:newcounterterm}
\end{align}
in terms of the arbitrary renormalization scale $M^2<4\mu^2$. Here the
vacuum polarization is written in term of vertices with with potential
$V$, and we have first carried out the
integrals over the loop energy $E$ and the Feynman parameter $\alpha$, and 
then shifted the integration variable to match
the branch cut along the imaginary axis, cf.\@ Sec. \ref{sec:interface}.
The expansion of the effective action,
Eq.~(\ref{eq:aeff2}), shows that this integral times
the average value
\begin{equation} 
\langle V^2(x) \rangle = \int_{-\infty}^{\infty} V^2(x) \, dx = 
\frac{4 \mu^3 (1+\ell)^2}{3 \ell}
\label{eq:V2}\end{equation}
gives the counterterm contribution to the energy (which is minus the spatial 
integral of the Lagrangian) subject to the condition 
that the renormalized polarization function vanishes at $p^2=M^2$.

Since the imaginary momentum integral in Eq.~(\ref{eq:newcounterterm}) is now of exactly the 
form as in Eq.~(\ref{eq:VPE2002}), we can write the full renormalized
energy per unit area as
\begin{equation}
\mathcal{E}_M^{(n)} = -\frac{1}{2 (4\pi)^{\frac{n+1}{2}}
\Gamma\left(\frac{n+3}{2}\right)} \int_\mu^\infty (t^2 - \mu^2)^{\frac{n+1}{2}}
\frac{\partial}{\partial t} \left[\nu(t) - \frac{\langle V \rangle}{2t}
+ \frac{\langle V^2 \rangle}{2t} \cdot \frac{1}{4t^2 - M^2} \right] \, dt \,.
\end{equation}
This expression is valid for any $n<2$. By incorporating the counterterm
directly into the momentum integral, we avoid the need to compute the
full Feynman diagram, requiring instead only the local quantities
$\langle V\rangle$ and $\langle V^2\rangle$. Most importantly, the
limit $n\,\to\,2$ is finite, giving
\begin{equation}
\mathcal{E}_M^{(2)}  = -\frac{1}{12\pi^2} \int_\mu^\infty
(t^2 - \mu^2)^{\frac{3}{2}} \frac{\partial}{\partial t} \left[\nu(t) + (\ell+1)\frac{\mu}{t}
+ \frac{2 \mu^3 (1+\ell)^2}{3 \ell t} \cdot \frac{1}{4 t^2 - M^2}
\right] \, dt \,.
\end{equation}
Carrying out the integral in both models, we obtain
\begin{align}
\mathcal{E}_{M,{\rm sG}}^{(2)} &=\frac{\mu^3}{6 \pi^2} 
 \left(\frac{2}{3} -  \frac{\sqrt{4 \mu^2 - M^2}}{M}
\arcsin\frac{M}{2 \mu}\right)\,,
\cr
\mathcal{E}_{M,{\rm kink}}^{(2)} &= \frac{3\mu^3}{16 \pi^2}
 \left(1 - \frac{\pi}{6\sqrt{3}}
- \frac{\sqrt{4 \mu^2 - M^2}}{M} \arcsin\frac{M}{2 \mu}
\right)
\end{align}
for the energy per unit area. For $M=0$, this result reproduces the VPE
presented in  Table~\ref{tab:2}. For the kink, $M=\mu$ agrees with the
$n=2$ result in the OS scheme, as shown in Table~\ref{tab:3}. For the
sine-Gordon solution, this calculation is academic because $V^2$ is not
part of the classical Lagrangian.

For consistency, we can also implement the same renormalization scheme
for the cases of $n=0$ and $n=1$.  In this case the second-order
counterterm simply makes a finite correction to the energy,
implementing the on-shell renormalization condition for the scattering
amplitude. These corrections are
\begin{equation}
\Delta \mathcal{E}_M^{(0)} = \frac{\mu}{\pi} \frac{(\ell+1)^2}{3\ell}
\frac{\mu^2}{M\sqrt{4 \mu^2 - M^2} } \arcsin\frac{M}{2\mu}
\end{equation}
for $n=0$ and
\begin{equation}
\Delta\mathcal{E}_M^{(1)} = \frac{\mu^2}{\pi} \frac{(\ell+1)^2}{12\ell} 
\frac{\mu}{M} \arctanh\frac{M}{2 \mu}
\end{equation}
for $n=1$. Adding $\Delta \mathcal{E}_M$ to the no-tadpole 
results from Eqs.~(\ref{eq:result0}) and~(\ref{eq:result1}) agrees
with the OS results in Table~\ref{tab:3}.

Similarly, we can also compute the finite correction obtained by
introducing a wavefunction renormalization counterterm $(\partial_\mu
\phi)^2$:  We expand the polarization function in a Taylor series
at the renormalization scale $M$ as
\begin{equation}
\Pi_V(p^2)=\Pi(M^2)+(p^2-M^2)\frac{\partial}{\partial M^2}\Pi_V(M^2)+\ldots \,,
\end{equation}
and set renormalization conditions to cancel the leading constant and
linear terms in this expansion.  The subtraction of the constant term
has already been implemented above.  For the linear term, the
counterterm to cancel the $p^2$ contribution contains 
$\phi'(x)^2$, while the counterterm to cancel the $M^2$ contribution
has $V(x)^2$, and the renormalization condition applies
only to the portion of the contribution quadratic in the deviation
of $\phi$ from its vacuum expectation value.

For the sine-Gordon model, the vertex interaction $U''(\phi)=-m^2 \cos\phi$ 
has no term linear in $\phi$, so wavefunction renormalization is absent in 
that case.  For the kink, the linear term in the vertex interaction  
$U''(\phi)=\frac{3\lambda}{2}(\phi^2-v^2)$ is $3 \mu\sqrt{\lambda} h$. 
For the soliton $h(x) = v\left(\tanh \frac{\mu x}{2}-1\right)$ we thus
have
\begin{equation}
\Delta\widetilde{\mathcal{E}}_M^{(n)} = 
\left(9 \mu^2\lambda \left\langle h'(x)^2 \right\rangle 
+  M^2 \left\langle V(x)^2 \right\rangle \right)
\frac{\partial}{\partial M^2} \Pi_V(M^2) \,.
\label{eqn:residueterm}
\end{equation}
The contribution from the $(\partial_\mu \phi)^2$ counterterm is
proportional to
\begin{equation}
\left\langle h'(x)^2 \right\rangle
= \int_{-\infty}^{\infty} h'(x)^2 \, dx  = \frac{2\mu^3}{3\lambda} \,,
\end{equation}
while the coefficient of the additional contribution proportional to
$\left\langle V(x)^2 \right\rangle$ is determined by the condition
that the combined correction to the inverse propagator that is
quadratic in $h$ should be proportional to $p^2-M^2$, with $V(x)
= 3\mu\sqrt{\lambda} h(x) + {\cal O}(h(x)^2)$.  Using
\begin{equation}
\frac{\partial }{\partial M^2}\Pi_V(M^2)=\frac{1}{2 (4\pi)^{\frac{n+1}{2}}
\Gamma\left(\frac{n+3}{2}\right)} \int_\mu^\infty (t^2 - \mu^2)^{\frac{n+1}{2}} 
\frac{\partial}{\partial t} \left[\frac{1}{2t} \frac{\partial}{\partial M^2}
\left(\frac{1}{4t^2 - M^2} \right)\right] \, dt \,,
\end{equation}
we obtain the additional contributions
\begin{align}
\Delta\widetilde{\mathcal{E}}_{M,{\rm kink}}^{(0)}&=
 \frac{3 \mu^3}{4 \pi M^3} (M^2+\mu^2) 
\frac{\frac{4 \mu^2-2M^2}{\sqrt{4 \mu^2-M^2}} 
\arcsin \frac{M}{2 \mu}-M} {4 \mu^2-M^2}\,,\cr
\Delta\widetilde{\mathcal{E}}_{M,{\rm kink}}^{(1)}&=
\frac{3\mu^3}{16 \pi  M^3} (M^2+\mu^2) 
\left({\rm artanh}\frac{M}{2\mu} -\frac{2 \mu M}{4 \mu^2-M^2}\right)\,,\cr
\Delta\widetilde{\mathcal{E}}_{M,{\rm kink}}^{(2)}&=
\frac{3\mu^3}{32 \pi ^2 M^3} (M^2+\mu^2) 
\left(M-\frac{4 \mu^2 \arcsin\frac{M}{2\mu}} {\sqrt{4\mu^2-M^2}}\right)\,.
\end{align}
For $M=\mu$ and $M\to 0$, all of these results agree with those
obtained numerically above for the ORS and ZM schemes, respectively.

\section{Conclusions}
\label{sec:conclusion}

Using spectral methods, we have analyzed the one-loop quantum 
corrections to the tensions of domain walls constructed from kink and sine-Gordon solitons. 
These are stationary solutions in one space dimension, which become domain 
walls when embedded in two or three dimensions. The whole configuration is 
translationally invariant in the additional coordinates and the
tension is the energy per unit length or area, respectively.

In spectral methods, the contribution of the continuum modes is computed from  scattering 
data for the interaction of the quantum fluctuations with the potential induced by the 
domain wall. A key feature of the approach is the equivalence between the Born expansion 
for scattering data and the expansion of the effective action in powers of the potential. 
The former is subtracted from the integrand in the continuum integral, and this subtraction 
is compensated for by adding back the latter at the corresponding order. The combination 
of the latter expansion with the counterterms renders the 
tension finite. For total dimensions $D<3+1$, it suffices to only consider the first order
of these expansions, which is simple. However, for $D=3+1$, a second-order subtraction is
necessary to produce finite results.  This subtraction can also be computed in $D=1+1$ 
and $D=2+1$, where it is finite and provides a check of our approach,
and also makes it possible to implement more general renormalization
schemes.  We can then extend all of these calculations to $D=3+1$ for the case
of the kink (the sine-Gordon model is not strictly renormalizable in
$D=3+1$). In all cases that we evaluated, the quantum correction for
the tension turned out to be negative.

For specific schemes, we have compared our calculations to results obtained
previously using different methods. While we agree with
Ref.~\cite{Rebhan:2002uk}, we observe large differences compared to
Ref.~\cite{Evslin:2024wwu}. We attribute this discrepancy to the
multiplicative  renormalization of the classical mass in
Ref.~\cite{Evslin:2024wwu} which does not incorporate the
renormalization of the ultraviolet divergence in the first order of
the expansions mentioned above.  More generally, we show how the
spectral approach provides a constructive, transparent, and
straightforward tool for the computation of the leading quantum contribution to tensions.

Finally, we have used dimensional regularization in the transverse
dimensions to directly implement all the necessary counterterms 
at an arbitrary energy scale within the integral over continuum
scattering modes. This approach avoids the need for explicit computation 
of the second-order contributions and produces analytic results.

\acknowledgments
N.\ G.\ is supported in part by the
National Science Foundation (NSF) through grant PHY-2209582.
H.\ W.\ is supported in part by the National Research Foundation of
South Africa (NRF) by grant~150672.

\subsection*{Note added}
After a preprint of this paper was made publicly available, a study 
appeared \cite{Evslin:2025krj} that addressed the discrepancy alluded to
in Table \ref{tab:evslin}. Those authors link that discrepancy
to the different points around which the fields are expanded in the
computation of Feynman diagrams in the perturbative sector. In our
approach, we adhere to the rule that the one-loop effective action is
to be computed by expanding around the classical  vacuum. The
ambiguity arising from the expansion about different vacua is absent
in commonly imposed renormalization schemes that leave the vacuum
expectation values of scalar fields unchanged. 

\bibliographystyle{apsrev}

\begin{thebibliography}{26}
\expandafter\ifx\csname natexlab\endcsname\relax\def\natexlab#1{#1}\fi
\expandafter\ifx\csname bibnamefont\endcsname\relax
  \def\bibnamefont#1{#1}\fi
\expandafter\ifx\csname bibfnamefont\endcsname\relax
  \def\bibfnamefont#1{#1}\fi
\expandafter\ifx\csname citenamefont\endcsname\relax
  \def\citenamefont#1{#1}\fi
\expandafter\ifx\csname url\endcsname\relax
  \def\url#1{\texttt{#1}}\fi
\expandafter\ifx\csname urlprefix\endcsname\relax\def\urlprefix{URL }\fi
\providecommand{\bibinfo}[2]{#2}
\providecommand{\eprint}[2][]{\url{#2}}

\bibitem[{\citenamefont{Rajaraman}(1982)}]{Ra82}
\bibinfo{author}{\bibfnamefont{R.}~\bibnamefont{Rajaraman}},
  \emph{\bibinfo{title}{Solitons and Instantons}} (\bibinfo{publisher}{North
  Holland}, \bibinfo{address}{Amsterdam}, \bibinfo{year}{1982}).

\bibitem[{\citenamefont{{Ya. B. Zeldovich, I. Yu. Kobzarev, and L. B.
  Okun}}(1974)}]{Zeldovich:1974uw}
\bibinfo{author}{\bibnamefont{{Ya. B. Zeldovich, I. Yu. Kobzarev, and L. B.
  Okun}}}, \bibinfo{journal}{Zh. Eksp. Teor. Fiz.}
  \textbf{\bibinfo{volume}{67}}, \bibinfo{pages}{3} (\bibinfo{year}{1974});
\bibinfo{journal}{Sov. Phys. JETP} \textbf{\bibinfo{volume}{40}}, 
\bibinfo{pages}{1} (\bibinfo{year}{1974}).

\bibitem[{\citenamefont{{T. W. B. Kibble}}(1976)}]{Kibble:1976sj}
\bibinfo{author}{\bibnamefont{{T. W. B. Kibble}}}, \bibinfo{journal}{J. Phys.
  A} \textbf{\bibinfo{volume}{9}}, \bibinfo{pages}{1387}
  (\bibinfo{year}{1976}).

\bibitem[{\citenamefont{{K. Aizu}}(1970)}]{PhysRevB.2.754}
\bibinfo{author}{\bibnamefont{{K. Aizu}}}, \bibinfo{journal}{Phys. Rev. B}
  \textbf{\bibinfo{volume}{2}}, \bibinfo{pages}{754} (\bibinfo{year}{1970}).

\bibitem[{\citenamefont{{G. F. Nataf, {\it et
  al.}}}(2020)}]{WOS:000574519100002}
\bibinfo{author}{\bibnamefont{{G. F. Nataf, {\it et al.}}}},
  \bibinfo{journal}{Nat. Rev. Phys.} \textbf{\bibinfo{volume}{2}},
  \bibinfo{pages}{634} (\bibinfo{year}{2020}).

\bibitem[{\citenamefont{{M. M. Salomaa and G. E.
  Volovik}}(1988)}]{Salomaa:1988zz}
\bibinfo{author}{\bibnamefont{{M. M. Salomaa and G. E. Volovik}}},
  \bibinfo{journal}{Phys. Rev. B} \textbf{\bibinfo{volume}{37}},
  \bibinfo{pages}{9298} (\bibinfo{year}{1988}).

\bibitem[{\citenamefont{{D. Tong}}(2005)}]{Tong:2005un}
\bibinfo{author}{\bibnamefont{{D. Tong}}}, in
  \emph{\bibinfo{booktitle}{{Theoretical Advanced Study Institute in Elementary
  Particle Physics}: {Many Dimensions of String Theory}}}
  (\bibinfo{year}{2005}), \eprint{hep-th/0509216}.

\bibitem[{\citenamefont{{A. Vilenkin}}(1985)}]{Vilenkin:1984ib}
\bibinfo{author}{\bibnamefont{{A. Vilenkin}}}, \bibinfo{journal}{Phys. Rept.}
  \textbf{\bibinfo{volume}{121}}, \bibinfo{pages}{263} (\bibinfo{year}{1985}).

\bibitem[{\citenamefont{{R. F, Dashen, B. Hasslacher, and A.
  Neveu}}(1974)}]{Dashen:1974cj}
\bibinfo{author}{\bibnamefont{{R. F, Dashen, B. Hasslacher, and A. Neveu}}},
  \bibinfo{journal}{Phys. Rev. D} \textbf{\bibinfo{volume}{10}},
  \bibinfo{pages}{4130} (\bibinfo{year}{1974}).

\bibitem[{\citenamefont{{A. Rebhan, P. van Nieuwenhuizen, and R.
  Wimmer}}(2002)}]{Rebhan:2002uk}
\bibinfo{author}{\bibnamefont{{A. Rebhan, P. van Nieuwenhuizen, and R. Wimmer}}},
  \bibinfo{journal}{New J. Phys.} \textbf{\bibinfo{volume}{4}},
  \bibinfo{pages}{31} (\bibinfo{year}{2002}).

\bibitem[{\citenamefont{{Evslin, Jarah} et~al.}(2025)\citenamefont{{Evslin,
  Jarah}, {Liu, Hui}, and {Zhang, Baiyang}}}]{Evslin:2024wwu}
\bibinfo{author}{\bibnamefont{{J. Evslin}}},
  \bibinfo{author}{\bibnamefont{{H. Liu}}}, \bibnamefont{and}
  \bibinfo{author}{\bibnamefont{{B. Zhang}}}, \bibinfo{journal}{Eur.
  Phys. J. C} \textbf{\bibinfo{volume}{85}}, \bibinfo{pages}{639}
  (\bibinfo{year}{2025}).

\bibitem[{\citenamefont{Graham et~al.}(2009)\citenamefont{Graham, Quandt, and
  Weigel}}]{Graham:2009zz}
\bibinfo{author}{\bibfnamefont{N.}~\bibnamefont{Graham}},
  \bibinfo{author}{\bibfnamefont{M.}~\bibnamefont{Quandt}}, \bibnamefont{and}
  \bibinfo{author}{\bibfnamefont{H.}~\bibnamefont{Weigel}},
  \emph{\bibinfo{title}{{Spectral Methods in Quantum Field Theory}}}, vol.
  \bibinfo{volume}{777, Lecture Notes Phys.}
  (\bibinfo{publisher}{Springer-Verlag, Berlin}, \bibinfo{year}{2009}).

\bibitem[{\citenamefont{Graham and Weigel}(2022)}]{Graham:2022rqk}
\bibinfo{author}{\bibfnamefont{N.}~\bibnamefont{Graham}} \bibnamefont{and}
  \bibinfo{author}{\bibfnamefont{H.}~\bibnamefont{Weigel}},
  \bibinfo{journal}{Int. J. Mod. Phys. A} \textbf{\bibinfo{volume}{37}},
  \bibinfo{pages}{2241004} (\bibinfo{year}{2022}).

\bibitem[{\citenamefont{{N. Graham, R. L. Jaffe, M. Quandt, and H.
  Weigel}}(2001{\natexlab{a}})}]{Graham:2001dy}
\bibinfo{author}{\bibnamefont{{N. Graham, R. L. Jaffe, M. Quandt, and H. Weigel}}},
  \bibinfo{journal}{Phys. Rev. Lett.} \textbf{\bibinfo{volume}{87}},
  \bibinfo{pages}{131601} (\bibinfo{year}{2001}{\natexlab{a}}).

\bibitem[{\citenamefont{{J. S. Faulkner}}(1977)}]{Faulkner:1977aa}
\bibinfo{author}{\bibnamefont{{J. S. Faulkner}}}, \bibinfo{journal}{J. Phys. C}
  \textbf{\bibinfo{volume}{10}}, \bibinfo{pages}{4661} (\bibinfo{year}{1977}).

\bibitem[{\citenamefont{{N. Graham, R. L. Jaffe, M. Quandt, and H.
  Weigel}}(2001{\natexlab{b}})}]{Graham:2001iv}
\bibinfo{author}{\bibnamefont{{N. Graham, R. L. Jaffe, M. Quandt, and H. Weigel}}},
  \bibinfo{journal}{Annals Phys.} \textbf{\bibinfo{volume}{293}},
  \bibinfo{pages}{240} (\bibinfo{year}{2001}{\natexlab{b}}).

\bibitem[{\citenamefont{{R. D. Puff}}(1975)}]{Puff:1975zz}
\bibinfo{author}{\bibnamefont{{R. D. Puff}}}, \bibinfo{journal}{Phys. Rev. A}
  \textbf{\bibinfo{volume}{11}}, \bibinfo{pages}{154} (\bibinfo{year}{1975}).

\bibitem[{\citenamefont{Newton}(1982)}]{Newton:1982qc}
\bibinfo{author}{\bibfnamefont{R.~G.} \bibnamefont{Newton}},
  \emph{\bibinfo{title}{Scattering Theory of Waves and Particles}}
  (\bibinfo{publisher}{Springer, New York}, \bibinfo{year}{1982}).

\bibitem[{\citenamefont{{M. Bordag}}(1995)}]{Bordag:1994jz}
\bibinfo{author}{\bibnamefont{{M. Bordag}}}, \bibinfo{journal}{J. Phys. A}
  \textbf{\bibinfo{volume}{28}}, \bibinfo{pages}{755} (\bibinfo{year}{1995}).

\bibitem[{\citenamefont{{S. Jaimungal, G. W. Semenoff, and K.
  Zarembo}}(1999)}]{Jaimungal:1998hk}
\bibinfo{author}{\bibnamefont{{S. Jaimungal, G. W. Semenoff, and K. Zarembo}}},
  \bibinfo{journal}{JETP Lett.} \textbf{\bibinfo{volume}{69}},
  \bibinfo{pages}{509} (\bibinfo{year}{1999}).

\bibitem[{\citenamefont{{N. Graham, R. L. Jaffe, V. Khemani, M. Quandt, M.
  Scandurra, and H. Weigel}}(2002)}]{Graham:2002xq}
\bibinfo{author}{\bibnamefont{{N. Graham, R. L. Jaffe, V. Khemani, M. Quandt,
  M. Scandurra, and H. Weigel}}}, \bibinfo{journal}{Nucl. Phys. B}
  \textbf{\bibinfo{volume}{645}}, \bibinfo{pages}{49} (\bibinfo{year}{2002}).

\bibitem[{\citenamefont{de~Vega}(1976)}]{deVega:1976sm}
\bibinfo{author}{\bibfnamefont{H.~J.} \bibnamefont{de~Vega}},
  \bibinfo{journal}{Nucl. Phys. B} \textbf{\bibinfo{volume}{115}},
  \bibinfo{pages}{411} (\bibinfo{year}{1976}).

\bibitem[{\citenamefont{{G. 't Hooft and M. J. G.
  Veltman}}(1979)}]{tHooft:1978jhc}
\bibinfo{author}{\bibnamefont{{G. 't Hooft and M. J. G. Veltman}}},
  \bibinfo{journal}{Nucl. Phys. B} \textbf{\bibinfo{volume}{153}},
  \bibinfo{pages}{365} (\bibinfo{year}{1979}).

\bibitem[{\citenamefont{Derrick}(1964)}]{Derrick:1964ww}
\bibinfo{author}{\bibfnamefont{G.~H.} \bibnamefont{Derrick}},
  \bibinfo{journal}{J. Math. Phys.} \textbf{\bibinfo{volume}{5}},
  \bibinfo{pages}{1252} (\bibinfo{year}{1964}).

\bibitem[{\citenamefont{Graham and Olum}(2003)}]{Graham:2002yr}
\bibinfo{author}{\bibfnamefont{N.}~\bibnamefont{Graham}} \bibnamefont{and}
  \bibinfo{author}{\bibfnamefont{K.~D.} \bibnamefont{Olum}},
  \bibinfo{journal}{Phys. Rev. D} \textbf{\bibinfo{volume}{67}},
  \bibinfo{pages}{085014} (\bibinfo{year}{2003});
  Erratum: Phys. Rev. D \textbf{69}, 109901 (2004)

\bibitem[{\citenamefont{{J. Evslin and H. Liu}}(2025)}]{Evslin:2025krj}
\bibinfo{author}{\bibnamefont{{J. Evslin and H. Liu}}} (\bibinfo{year}{2025}),
  \eprint{2505.21856}.

\end{thebibliography}

\end{document}